\documentclass[
reprint,
longbibliography,
amsmath,amssymb,
aps,
prl,
floatfix,
]{revtex4-2}

\usepackage{graphicx}%
\usepackage[dvipsnames]{xcolor}
\usepackage{cancel}
\usepackage{dcolumn}%
\usepackage{bm}%
\usepackage[colorlinks, linkcolor=blue, citecolor=blue, urlcolor=blue]{hyperref}%
\usepackage{booktabs}

\begin{document}

\title{A Bound on Topological Gap from Newton's Laws}
\author{Navketan Batra}
\author{D. E. Feldman}
\affiliation{Department of Physics, Brown University, Providence, Rhode Island 02912, USA}
\affiliation{Brown Theoretical Physics Center, Brown University, Providence, Rhode Island 02912, USA}

\date{\today}

\begin{abstract}
A striking general bound on the energy gap in topological matter was recently discovered in Ref. [Onishi and Fu, Phys. Rev. X {\bf 14}, 011052 (2024)]. A non-trivial indirect derivation builds on the properties of optical conductivity at an arbitrary frequency. 
We propose a simpler derivation, 
{\color{black} allowing} multiple generalizations, such as a universal bound on a gap in anisotropic systems, systems with multiple charge carrier types, and  topological systems with zero Hall conductance. The derivation builds on the observation that the bound equals $\hbar$ times the ratio of the force by the external electric field on the charge carriers and their total kinematic momentum in the direction perpendicular to the force.

\end{abstract}

\maketitle

Gapped systems cannot absorb energy at the frequencies below ${\rm gap}/\hbar$. Hence, their longitudinal d.c. conductance is zero. On the other hand, a nonzero quantized Hall conductance is allowed  in topological insulators \cite{WenBook,book}. The Hall conductance $\sigma_{xy}=Ce^2/h$ with an integer Chern number $C$ in non-interacting \textcolor{black}{two dimensional (2D)} electronic systems. Fractional \textcolor{black}{values of} $C$ are allowed in the presence of electron interaction. At first sight,  Hall conductance tells nothing about the energy gap. Indeed, one can build a model with an {\color{black} arbitrarily} large gap and zero $C$: just confine each electron in a harmonic trap with an {\color{black} arbitrarily} large level spacing. Yet, in an unexpected twist, Onishi and Fu discovered
\cite{OF,jclub} that a nonzero $C$ puts a universal upper bound on the {\color{black}optical} energy gap $\epsilon_{\rm gap}$.

The bound is remarkably simple:
\begin{equation}
\label{1}
\epsilon_{\rm gap}\le \frac{2\pi\hbar^2 n}{|C|m},
\end{equation}
where $m$ is an electron mass and $n$ is the electron density in 2D.
The derivation is{\color{black},} however{\color{black},} rather nontrivial and involves properties of the response to high-frequency circularly polarized light. The goal of this paper is to offer a more intuitive{\color{black},} rigorous derivation. As we will see, a simpler derivation opens a way to multiple generalizations of the result, including even to some situations with zero $\sigma_{xy}$ in two or more dimensions \cite{bernevig2006, konig2007, wu2018, li2021, kang2024}. The idea comes from the following elementary observation.

Apply a uniform electric field $E_x$ along the $x$-axis in a 2D system. The force acting on the electrons
\begin{equation}
\label{Fx}
F_x=eE_xnS, 
\end{equation}
where $e$ is an electron charge and $S$ is the area.
The current density $j_y=(Ce^2/h)E_x=nv_ye$, where $v_y$ is the average velocity. 
Hence, the kinematic momentum 
\begin{equation}
\label{Py}
P_y=Snmv_y=CemE_xS/h.
\end{equation}
Eq. (\ref{1}) can now be recast as
\begin{equation}
\label{2}
\epsilon_{\rm gap}\le \hbar\left|\frac{F_x}{P_y}\right|.
\end{equation}
This has a simple physical meaning. The average force acting on the electrons in a stationary state is 0. If we suddenly turn off the external field $E_x$, they will experience an internal force $-F_x$. This will cause rotation of the momentum with the frequency $\omega=F_x/P_y$. The existence of such characteristic frequency suggests 
the presence of energies $\epsilon\le \hbar|\omega|$ in the spectrum.
{\color{black} Indeed, the characterstic response time is set by the intrinsic time scales on the order of
$\hbar/\epsilon$, where $\epsilon$ is an excitation energy.}

We will now turn this heuristic into a rigorous argument.
As in Ref. \onlinecite{OF}, we assume a kinetic energy operator, quadratic in momentum for each particle,
\begin{equation}
\label{3}
\hat T=\frac{(\hat{\bf p}+\textcolor{black}{\hat{{\bf a}}}({\bf r}))^2}{2m},
\end{equation}
where ${\hat{\bf a}}({\bf r})$ is allowed to depend on spin. Such structure ensures the validity of the Ehrenfest theorem 
\cite{LL} for the average force
and the time derivative of the kinematic momentum. The Ehrenfest theorem is 
quantum version of Newton's second law.
The total Hamiltonian $H_0$ also includes inter-electron interaction and the interaction with the external periodic and random potentials. 

{\color{black} Besides the structure of the kinetic energy (\ref{3}), the Ehrenfest theorem
relies on the canonical commutation relations for the coordinate and dynamical-momentum operators.
The operators, acting on different particles, must commute, the components of the momentum must commute, the coordinates must commute, and the only nonzero commutators for each particle are 
$[\hat{x},\hat{p}_x]=[\hat{y},\hat{p}_y]=i\hbar$. These assumptions clearly hold if $\hat{\bf p}$ and
$\hat{\bf r}$ describe electrons in a microscopic model. In that case, $\hat{\bf a}({\bf r})$ includes
the vector potential of the external magnetic field, and the potential energy may be spin-dependent through Zeeman terms. The spin-orbit interaction $\hat{\bf p}\cdot (\hbar/4m^2c^2) [\nabla V \times 
\hat{\bf{\sigma}}]$ can be absorbed into $\hat{\bf a}$ as a spin-dependent contribution.
In a microscopic model, $m$ is the mass of an electron in vacuum and the concentration $n$ includes all electrons in the material.

It may also be possible to derive a bound on the optical gap from an effective low-energy model, so that $m$
becomes an effective mass and $n$ includes only electrons in topological bands. The bound only works if the kinetic energy is parabolic in a good approximation and the canonical commutation relations hold at least approximately in the low-energy model, obtained by integrating out high-energy degrees of freedom. As will be obvious from the derivation below, the integration-out procedure should retain all topological bands responsible for a non-zero $C$. It is well known that the quantum Hall effect in GaAs is well described 
by a model of electrons or holes populating a parabolic band and obeying the canonical commutation relations \cite{PG}. As discussed in Ref. {\onlinecite{OF}}, the anomalous quantum Hall effect in twisted MoTe$_2$
also derives from a model \cite{WLTMM} with parabolic dispersion and spin-dependent $\hat{\bf a}$
where our assumptions apply.
}

{\color{black} As in Ref. \onlinecite{OF}, we consider an infinite system. Indeed, 
a finite system has gapless edge states and hence no energy gap. Thus, the area $S$ is infinite. One can avoid the factor $S$ in equations by rewriting Eqs. (\ref{Fx})
and (\ref{Py}) in terms of the force and momentum densities. One can also eliminate $S$ below
by using the sum of all electron coordinates in place of the product of $S$, $n$, and the center-of-mass position.}

{\color{black} We will follow the above heuristic argument and consider the effect of a sudden switch of the electric field. It is possible to give a closely related argument without such a trick (see {\color{black}Appendix}).}
{\color{black} We denote the energy of an excited  state $|k\rangle$
as $\epsilon_k>0${\color{black}}
and, without loss of generality{\color{black},} set the energy of the ground state $|0\rangle$ to 0.}
In the presence of a weak electric field $E_x${\color{black},} the Hamiltonian gets a correction $-eE_xnS\hat{X}$, where $X$ is the center-of-mass position. In the first-order perturbation theory, the ground state becomes
\begin{equation}
\label{4}
|\psi_x\rangle=|0\rangle+eE_xnS\sum_{k\ne 0}\frac{|k\rangle\langle k|\hat{X}|0\rangle}{\epsilon_k}.
\end{equation}
When $E_x$ is suddenly turned off, the state does not change immediately, and the internal force remains
\begin{equation}
\label{5}
-F_x=\langle \psi_x|\frac{dP_x}{dt}|\psi_x\rangle,
\end{equation}
where $F_x$ is given by Eq. (\ref{Fx}), the time derivative of the kinematic momentum ${\color{black}d\hat{P}_x/dt}=i[H_0,P_x]/\hbar$, and the kinematic momentum 
$\hat{P}_x=mSn{\color{black}d\hat{X}/dt}=imSn[H_0,\hat{X}]/\hbar$. A substitution of this definition into Eq. (\ref{5}) yields
\begin{equation}
\label{6}
F_x=\frac{2en^2mS^2E_x}{\hbar^2}\sum_{k\ne 0}\epsilon_k|\langle0|\hat{X}|k\rangle|^2.
\end{equation}
A similar calculation of the average kinematic momentum in the $y$-direction 
$\hat{P}_y=Snm{\color{black}d\hat Y/dt}$ yields
\begin{equation}
\label{7}
P_y=\frac{ieE_xn^2S^2m}{\hbar}\sum_{k\ne 0}[\langle 0|\hat{X}|k\rangle\langle k|\hat{Y}|0\rangle-\langle 0|\hat{Y}|k\rangle\langle k|\hat{X}|0\rangle].
\end{equation}
The ratio of the left-hand sides of Eqs. (\ref{6}) and (\ref{7}) is $F_x/P_y=hn/Cm=\omega$. 
{\color{black} This gives us the relation
\begin{equation}
\label{6b}
2\sum_{k\ne 0}\epsilon_k|\langle0|\hat{X}|k\rangle|^2 = i\hbar \omega \sum_{k\ne 0}[\langle 0|\hat{X}|k\rangle\langle k|\hat{Y}|0\rangle-\text{c.c.}].
\end{equation}}
\textcolor{black}{where c.c. denote complex conjugate.}

The Hall current and the electric force just change their directions and preserve their magnitudes, if the electric field is applied along the $y$-axis. Therefore, analogous relations can be found by considering the effect of the electric field along the $y$-direction. The ratio of the internal force and the kinematic momentum is now $F_y/P_x=-hn/Cm=-\omega$, which yields,
\begin{equation}
\label{6a}
2\sum_{k\ne 0}\epsilon_k|\langle0|\hat{Y}|k\rangle|^2 = i\hbar \omega \sum_{k\ne 0}[\langle 0|\hat{X}|k\rangle\langle k|\hat{Y}|0\rangle-\text{c.c.}].
\end{equation}
By combining Eqs.  (\ref{6b}) and (\ref{6a}) we arrive at,
\begin{eqnarray}
\label{8}
\sum_{k\ne 0}\epsilon_k(|\langle 0|\hat{Y}|k\rangle|^2+|\langle 0|\hat{X}|k\rangle|^2 )=
\nonumber & &\\
i\hbar\omega\sum_{k\ne 0}[\langle 0|\hat{X}|k\rangle\langle k|\hat{Y}|0\rangle-\langle 0|\hat{Y}|k\rangle\langle k|\hat{X}|0\rangle] & & \nonumber\\
\le \hbar|\omega|\sum_{k\ne 0}2|\langle 0|\hat{X}|k\rangle\langle k|\hat{Y}|0\rangle|. & &
\end{eqnarray}
Finally, we apply  the inequality
$| \langle 0|\hat{X}|k\rangle\langle k|\hat{Y}|0\rangle|\le (|\langle 0|\hat{Y}|k\rangle|^2+|\langle 0|\hat{X}|k\rangle|^2 )/2 $ and obtain
\begin{equation}
\label{9}
\sum_{k\ne 0}(\epsilon_k-\hbar|\omega|)(|\langle 0|\hat{Y}|k\rangle|^2+|\langle 0|\hat{X}|k\rangle|^2 )\le 0.
\end{equation}
The bound (\ref{1},\ref{2}) follows since the left-hand side is positive when all $\epsilon_k>\hbar|\omega|$.

{\color{black} In a modified argument, one places the system on a torus to eliminate gapless edges. The argument remains essentially the same with two modifications. First, the electric field is introduced through a time-dependent low-frequency 
magnetic flux, parallel to the surface of the torus. Second, one focuses on the momentum of the particles in place of the time derivatives of multi-valued coordinates. {\color{black} The resulting proof is shorter and more rigorous  though perhaps less intuitive than the above argument (see {\color{black}Appendix}).}}

Our derivation makes no assumptions about neutral degrees of freedom. A topological gap can be affected by interaction with {\color{black} the} electromagnetic field in a resonator \cite{Res}. The bound still must hold. The derivation also easily accommodates anisotropy when the effective mass depends on the direction, that is, the quadratic in the momentum part of the energy of each particle is $p_x^2/2m_x+p_y^2/2m_y$. 
Two different masses {\color{black} now enter} the definitions of the kinematic momenta in two directions,
$\hat{P_{x}}=Snm_{x}d\hat{X}/dt$ and $\hat{P_{y}}=Snm_{y}d\hat{Y}/dt$.
The same steps lead to the bound
\begin{equation}
\label{10}
\epsilon_{\rm gap}\le \frac{\pi \hbar^2 n(m_x+m_y)}{|C|m_xm_y}.
\end{equation}

The above argument makes  minimal use of two-dimensionality, and it is straightforward to extend the bound to insulators in \textcolor{black}{three dimensions (3D)}. 
{\color{black} We just introduce the 3D charge density $n_{\text{3D}}$ {\color{black} so that the force per unit volume} $F_x=en_{\text{3D}}E_x$. 
The Hall current density $j_y=\sigma_{xy}E_x=n_{\text{3D}}v_y e$. The same derivation yields
\begin{equation}
\label{3D}
\epsilon_{\rm gap}\le \frac{\hbar e^2 n_{\text{3D}}}{|\sigma_{xy}| m}.
\end{equation}
}

Another generalization is to systems with multiple carrier types with different charges $e_\alpha$ and masses $m_\alpha$. This is relevant, in particular, for a low-energy effective theory in which carriers with different effective masses are present. For example, this may happen in a van der Waals structure with layers of different nature{\color{black}s}. Bilayer topological liquids are well established experimentally \cite{bilayer}.

When we turn off a weak electric field $E_x$ along the $x$-axis, the total force on carriers of type $\alpha$ becomes $-F_{x,\alpha}={\color{black} dP_{x,\alpha}/dt}=-e_\alpha E_x n_\alpha S$, where $n_\alpha$ is their density and $P_{x,\alpha}$ is the $x$-component of the kinematic momentum. This can be used to find the second derivative of the dipole moment $D_x=\sum_\alpha Sn_\alpha e_\alpha X_\alpha$, where $X_\alpha$ is the center-of-mass coordinate for carriers of type $\alpha$. We get
\begin{equation}
\label{11}
{\color{black}\frac{d^2}{dt^2}} D_x=\sum_\alpha \frac{e_\alpha}{m_\alpha}{\color{black}\frac{d}{dt}}P_{x,\alpha}=-E_xS\sum_\alpha \frac{e_\alpha^2n_\alpha}{m_\alpha}.
\end{equation}
We focus on the dipole moment because it couples to the electric field $E_x${\color{black},} so the field-dependent correction to the Hamiltonian is $-E_xD_x$. The perturbed state $|\psi_x\rangle$ is {\color{black} now given} by a modified Eq. (\ref{4}) with ${\hat{D}}_x/e{\color{black}Sn}$ in place of $\hat X$. Similar to Eq. (\ref{5}) we equate the expression (\ref{11}) and the average of the operator ${\color{black} d^2{{\hat{D}}_x}/dt^2}$ in the state $|\psi_x\rangle$. We find
\begin{equation}
\label{12}
{\color{black}\frac{d^2}{dt^2}} D_x=-\frac{2 E_x}{\hbar^2}\sum_{k\ne 0}\epsilon_k|\langle 0|\hat{D}_x|k\rangle|^2.
\end{equation}
We now turn to the electric current density
\begin{equation}
\label{13}
j_y=\sigma_{xy}E_x.
\end{equation}
It can also be computed as the average of the operator $\hat{j}_y=\sum_{\alpha}n_\alpha e_\alpha {\color{black} d{\hat{Y}}_\alpha/dt}={\color{black} S^{-1} d {\hat{D}}_y/dt}$ in the state $|\psi_x\rangle$,
\begin{equation}
\label{14}
j_y=\frac{iE_x}{\hbar{\color{black}S}}\sum_{k\ne 0}[\langle0|\hat{D}_x|k\rangle\langle k|\hat{D}_y|0\rangle-
\langle0|\hat{D}_y|k\rangle\langle k|\hat{D}_x|0\rangle].
\end{equation}
The rest of the argument is essentially the same as for a single carrier type. First, by considering the effect of the electric field {\color{black}$E_x$} along the $y$-direction, we get an {\color{black} analog} of Eq. (\ref{12}) for matrix elements of $\hat{D}_y$,
\begin{equation}
\label{15}
-E_xS\sum_\alpha \frac{e_\alpha^2n_\alpha}{m_\alpha}=-\frac{2 E_x}{\hbar^2}\sum_{k\ne 0}\epsilon_k|\langle 0|\hat{D}_y|k\rangle|^2.
\end{equation}
Next,
combining Eqs. (\ref{11},\ref{12},\ref{13},\ref{14},\ref{15}) we find that
\begin{eqnarray}
\label{16}
\sum_{k\ne 0}\epsilon_k(|\langle 0|\hat{D}_x|k\rangle|^2+|\langle 0|\hat{D}_y|k\rangle|^2)=~~~~~~& &\nonumber\\
\frac{i\hbar}{\sigma_{xy}}\sum_\alpha\frac{e_\alpha^2 n_\alpha}{m_\alpha}
\sum_{k\ne 0}[\langle 0|\hat{D}_x|k\rangle\langle k|\hat{D}_y|0\rangle-
\langle 0|\hat{D}_y|k\rangle\langle k|\hat{D}_x|0\rangle]& &\nonumber\\
\le \frac{2\hbar}{|\sigma_{xy}|}\sum_\alpha\frac{e_\alpha^2 n_\alpha}{m_\alpha}
\sum_{k\ne 0}|\langle 0|\hat{D}_x|k\rangle\langle k|\hat{D}_y|0\rangle|~~~~~~& &\nonumber\\
\le \frac{\hbar}{|\sigma_{xy}|}\sum_\alpha\frac{e_\alpha^2 n_\alpha}{m_\alpha}
\sum_{k\ne 0}(|\langle 0|\hat{D}_x|k\rangle|^2+|\langle 0|\hat{D}_y|k\rangle|^2).~~~~~~~ & &
\end{eqnarray}
The inequality can only hold if at least one energy level satisfies the bound
\begin{equation}
\label{17}
\epsilon_k\le\frac{\hbar}{|\sigma_{xy}|}\sum_\alpha\frac{e_\alpha^2 n_\alpha}{m_\alpha}.
\end{equation}

We finally turn to a situation, where an electric field does not cause a Hall current or a nonzero momentum in the perpendicular direction. As an example, we consider a spin Hall system \cite{SHE} in which a spin current
$j_s=CeE_x/4\pi$ is present due to the opposite electric currents of spin-up and -down electrons
\begin{equation}
\label{18}
j_{\uparrow, y}=-j_{\downarrow, y}=\frac{Ce^2E_x}{2h}.
\end{equation}
{\color{black}We assume no spin-flip processes so  the Ehrenfest theorem applies separately to spin-up and -down electrons.} {\color{black} We also assume equal densitites $n/2$ of spin-up and -down particles.}

{\color{black} We will consider fictitious electric fields ${\bf E}_{\uparrow}=(E_{\uparrow,x}, E_{\uparrow,y})$ and ${\bf E}_{\downarrow}=(E_{\downarrow,x}, E_{\downarrow,y})$, which are coupled to only spin-up or only spin-down electrons.
The physical electric field ${\bf E}$ corresponds to ${\bf E}={\bf E}_\uparrow={\bf E}_\downarrow$. In linear response,

\begin{equation}
\label{matrix_current_up}
\begin{pmatrix}
j_{\uparrow,x}\\
j_{\uparrow,y}
\end{pmatrix}
=
\begin{pmatrix}
\sigma_{\uparrow\uparrow} & -\frac{C_{\uparrow\uparrow}}{2}\\
\frac{C_{\uparrow\uparrow}}{2} & \sigma_{\uparrow\uparrow}
\end{pmatrix}
\begin{pmatrix}
E_{\uparrow,x}\\
E_{\uparrow,y}
\end{pmatrix}
+
\begin{pmatrix}
\sigma_{\downarrow\uparrow} & -\frac{C_{\downarrow\uparrow}}{2}\\
\frac{C_{\downarrow\uparrow}}{2} & \sigma_{\downarrow\uparrow}
\end{pmatrix}
\begin{pmatrix}
E_{\downarrow,x}\\
E_{\downarrow,y}
\end{pmatrix}
\end{equation}
and

\begin{equation}
\label{matrix_current_down}
\begin{pmatrix}
j_{\downarrow,x}\\
j_{\downarrow,y}
\end{pmatrix}
=
\begin{pmatrix}
\sigma_{\uparrow\downarrow} & -\frac{C_{\uparrow\downarrow}}{2}\\
\frac{C_{\uparrow\downarrow}}{2} & \sigma_{\uparrow\downarrow}
\end{pmatrix}
\begin{pmatrix}
E_{\uparrow,x}\\
E_{\uparrow,y}
\end{pmatrix}
+
\begin{pmatrix}
\sigma_{\downarrow\downarrow} & -\frac{C_{\downarrow\downarrow}}{2}\\
\frac{C_{\downarrow\downarrow}}{2} & \sigma_{\downarrow\downarrow}
\end{pmatrix}
\begin{pmatrix}
E_{\downarrow,x}\\
E_{\downarrow,y}
\end{pmatrix},
\end{equation}
where the diagonal and non-diagonal matrix elements stay for the longitudinal and Hall
conductances. The response to the physical electric field ${\bf E}={\bf E}_\uparrow={\bf E}_\downarrow$ imposes the constraints $\sigma_{\uparrow\uparrow}=-\sigma_{\downarrow\uparrow}$, $\sigma_{\downarrow\uparrow}=-\sigma_{\downarrow\downarrow}$,
$C_{\uparrow\uparrow}+C_{\downarrow\uparrow}=Ce^2/h$, and $C_{\uparrow\downarrow}+C_{\downarrow\downarrow}=-Ce^2/h$.

Since the system is gapped, there is no dissipative transport, that is,
${\bf E}_\uparrow\cdot {\bf j}_\uparrow+{\bf E}_\downarrow\cdot {\bf j}_\downarrow=0$.
This gives additional constraints $\sigma_{\rm any~indexes}=0$ and $C_{\uparrow\uparrow}-
C_{\downarrow\downarrow}=2Ce^2/h$. Thus, linear response simplifies to

\begin{align}
\label{matrix_current_up_1}
\begin{pmatrix}
j_{\uparrow,x}\\
j_{\uparrow,y}
\end{pmatrix}
&=
\begin{pmatrix}
0 & -\frac{C_{\uparrow\uparrow}}{2}\\
\frac{C_{\uparrow\uparrow}}{2} & 0
\end{pmatrix}
\begin{pmatrix}
E_{\uparrow,x}\\
E_{\uparrow,y}
\end{pmatrix}
\nonumber \\
&~~+
\begin{pmatrix}
0 & \frac{C_{\uparrow\uparrow}-{\color{black}G_0}C}{2}\\
\frac{{\color{black}G_0}C-C_{\uparrow\uparrow}}{2} & 0
\end{pmatrix}
\begin{pmatrix}
E_{\downarrow,x}\\
E_{\downarrow,y}
\end{pmatrix}
\end{align}
and
\begin{eqnarray}
\label{matrix_current_down_1}
\begin{pmatrix}
j_{\downarrow,x}\\
j_{\downarrow,y}
\end{pmatrix}
=
\begin{pmatrix}
0 & {\color{black}\frac{C_{\uparrow\uparrow}-{\color{black}G_0}C}{2}}\\
\frac{{\color{black}G_0}C-C_{\uparrow\uparrow}}{2} & 0
\end{pmatrix}
\begin{pmatrix}
E_{\uparrow,x}\\
E_{\uparrow,y}
\end{pmatrix}
& & \nonumber\\
+
\begin{pmatrix}
0 & \frac{2{\color{black}G_0}C-C_{\uparrow\uparrow}}{2}\\
\frac{C_{\uparrow\uparrow}-2{\color{black}G_0}C}{2} & 0
\end{pmatrix}
\begin{pmatrix}
E_{\downarrow,x}\\
E_{\downarrow,y}
\end{pmatrix}, & &
\end{eqnarray}
where $G_0=e^2/h$.
We can now estimate the gap in two ways: 1) by looking at the response of spin-up particles to the field
${\bf E}_\uparrow$ at zero ${\bf E}_\downarrow$; 2) by looking at the response of spin-down particles to the field
${\bf E}_\downarrow$ at zero ${\bf E}_\uparrow$.
The two estimates are most easily obtained with our initial classical-mechanical argument from the ratios of the electric forces acting on spin-up$/$down electrons to their total kinematic momenta. 
The two upper bounds are ${\hbar e^2 n}/{|C_{\uparrow\uparrow}|m}$ and 
${\hbar e^2 n}/{|2CG_0-C_{\uparrow\uparrow}|m}$. We observe that ${\rm max}(|C_{\uparrow\uparrow}|,
|2CG_0-C_{\uparrow\uparrow}|)\ge |CG_0|$. Finally,

\begin{equation}
\epsilon_{\rm gap}\le\frac{2\pi\hbar^2 n}{|C|m}.
\end{equation}
}

Besides charge and spin currents, much interest has been attracted to quantized heat currents in topological matter \cite{review, KF, N1, N2}. It is unclear if our approach extends to heat currents. Indeed, our derivation assumes that the current flows in the gapped bulk.
Yet, the low-temperature heat current in a gapped system flows along its edges. This is different from the electric current. Even though charge transport in the quantum Hall effect is usually {\color{black}modeled} with the one-dimensional chiral Luttinger model \cite{WenBook}, the current actually flows in the bulk of an incompressible system, and only the dynamic degrees of freedom live at the edges. 

In conclusion, we find a simple derivation of a large family of bounds on the energy gap in topological systems. The bounds apply to a generic situation when an external force, acting in the bulk of the system, causes a current in the bulk. 
We derive bounds for systems with topological electrical and spin currents {\color{black} with and without an external magnetic field \cite{cai2023,zeng2023,park2023,lu2024}}. The same derivation would apply to other types of transport, such as valleytronics \cite{val}.

We thank L. Fu for a stimulating discussion and G. E. Volovik for useful comments. This research was
supported in part by the National Science Foundation under Grant No. DMR-2204635 and
by grant NSF PHY-2309135 to the Kavli Institute for Theoretical Physics (KITP).
\vskip .3in

{\color{black}
\appendix
\section{Appendix: A proof on a torus}

A shorter and more rigorous proof is obtained by placing the system on a torus. The coordinate $x$ now runs along the great circles of the torus and $y$ runs along the orthogonal circles. The electric field is introduced 
with a vector potential $A=A_0\cos(\omega t)$ along the $x$ direction. The low-frequency uniform electric field $E=-dA/cdt$ points along the great circles. In place of the coordinates we use the operators of the average current densities along the $x$ and $y$ directions, $J_x-\frac{ne^2A}{mc}$ and $J_y$, where we explicitly show the $A$-dependent term in the definition of the current.

The Hall conductance is given by the Kubo formula,

\begin{equation}
\label{Kubo}
\frac{Ce^2}{h}=i S\hbar\sum_k\frac{\langle 0|J_x|k\rangle\langle k|J_y|0\rangle-\langle 0|J_y|k\rangle\langle k|J_x|0\rangle}{\epsilon^2_k},
\end{equation}
where $S$ is the area of the torus, $|0\rangle$ the ground state, 
and $\epsilon_k$ the energies of the excited states as before. Fractional quantum Hall systems have multiple ground states on a torus. This does not matter for our argument since nonzero matrix elements 
of current density as well as any other local operator only exist for states in the same topological sector, and the gap should be understood as the gap in a given topological sector.

The low-frequency longitudinal conductance is zero in an insulator. Yet, the calculation of the average 
current in the $x$ direction with the time-dependent perturbation theory produces a contribution of order $1/\omega$, where $\omega\rightarrow 0$. Clearly, the coefficient in front of $1/\omega$ must be $0$. This yields the equation
\begin{equation}
\label{f-rule}
\frac{ne^2}{m}=2S\sum_k\frac{|\langle 0|J_x|k\rangle|^2}{\epsilon_k}.
\end{equation}
By applying an electric field in the $y$ direction we get a similar result for $J_y$: 
\begin{equation}
\label{f-rule-2}
\frac{ne^2}{m}=2S\sum_k\frac{|\langle 0|J_y|k\rangle|^2}{\epsilon_k}.
\end{equation}
Finally, we apply the inequality of the arithmetic and geometric means, 
$ |\langle 0|J_x|k\rangle\langle k|J_y|0\rangle-\langle 0|J_y|k\rangle\langle k|J_x|0\rangle|
\le |\langle 0|J_x|k\rangle|^2+|\langle 0|J_y|k\rangle|^2$,
to Eq. (\ref{Kubo}). We find
\begin{eqnarray}
\label{ineq-final}
\frac{|C|e^2}{h}\le S\hbar\sum_k\frac{|\langle 0|J_x|k\rangle|^2+|\langle 0|J_y|k\rangle|^2}{\epsilon^2_k}
& & \nonumber\\
\le\frac{S\hbar}{\epsilon_{\rm gap}}\sum_k\frac{|\langle 0|J_x|k\rangle|^2+|\langle 0|J_y|k\rangle|^2}{\epsilon_k} =\frac{ne^2\hbar}{m\epsilon_{\rm gap}},
& &
\end{eqnarray}
where we used Eqs. (\ref{f-rule}) and (\ref{f-rule-2}) at the last step. The bound on the optical gap follows.}

\end{document}